\documentclass{iaus}
\usepackage{graphicx}

\title[Observing an Event Horizon with mm VLBI]{Observing a Black Hole Event
  Horizon: \\ (Sub)Millimeter VLBI of Sgr~A*}

\author[Vincent L.\ Fish \& Sheperd S.\ Doeleman]{Vincent L.\ Fish \& Sheperd S.\ Doeleman}

\affiliation{MIT Haystack Observatory \\
  Off Route 40\\
  Westford, MA  01886, USA\\
  email: {\tt vfish@haystack.mit.edu, sdoeleman@haystack.mit.edu}}

\pubyear{2009}
\volume{261}
\pagerange{xxx--yyy}
\jname{Relativity in Fundamental Astronomy:\\ 
  Dynamics, Reference Frames, and Data Analysis}
\editors{S. Klioner, P.K. Seidelmann, \& M. Soffel, eds.}
\begin{document}

\maketitle

\begin{abstract}
Very strong evidence suggests that Sagittarius~A*, a compact radio
source at the center of the Milky Way, marks the position of a super
massive black hole.  The proximity of Sgr~A* in combination with its
mass makes its apparent event horizon the largest of any black hole
candidate in the universe and presents us with a unique opportunity to
observe strong-field GR effects.  Recent millimeter very long baseline
interferometric observations of Sgr~A* have demonstrated the existence
of structures on scales comparable to the Schwarzschild radius.  These
observations already provide strong evidence in support of the
existence of an event horizon.  (Sub)Millimeter VLBI observations in
the near future will combine the angular resolution necessary to
identify the overall morphology of quiescent emission, such as an
accretion disk or outflow, with a fine enough time resolution to
detect possible periodicity in the variable component of emission.  In
the next few years, it may be possible to identify the spin of the
black hole in Sgr~A*, either by detecting the periodic signature of
hot spots at the innermost stable circular orbit or parameter
estimation in models of the quiescent emission.  Longer term, a
(sub)millimeter VLBI ``Event Horizon Telescope'' will be able to
produce images of the Galactic center emission to the see the
silhouette predicted by general relativistic lensing.  These
techniques are also applicable to the black hole in M87, where black
hole spin may be key to understanding the jet-launching region.

%% add here a maximum of 10 keywords, to be taken form the file <Keywords.txt>
\end{abstract}

\firstsection
\section{Introduction}

The Galactic center radio source Sagittarius~A* is believed to host a
massive ($\sim 4 \times 10^6$~M$_\mathrm{sun}$) black hole.  Due to
its proximity at $\sim 8$~kpc (e.g., \cite[Reid 2008]{reid08}),
Sgr~A* has the largest apparent event horizon of any known black hole
candidate: $r_\mathrm{Sch} = 2 r_\mathrm{G} \approx 10~\mu$as.

There is active debate in the scientific community on whether the
emission from Sgr~A* is predominantly due to an accretion disk or a
jet.  Further data are necessary in order to disentangle the
accretion/outflow physics from general relativistic effects.

There are several reasons to observe Sgr~A* at millimeter wavelengths.
The spectrum of Sgr~A* peaks in the millimeter.  Interstellar
scattering, which varies as $\lambda^2$, dominates over intrinsic
source structure at longer wavelengths, and the emission from Sgr~A*
transitions from optically thick to optically thin near $\lambda =
1$~mm (\cite[Doeleman \etal\ 2001]{doeleman01}).  Finally, millimeter
wavelengths permit observations by the technique of very long baseline
interferometry (VLBI).  The spatial scales accessible to millimeter
VLBI range down to a few $r_\mathrm{G}$, providing angular resolution
currently unachievable by any other means.

Recent VLBI observations by \cite[Doeleman \etal\ (2008)]{doeleman08}
were successful in detecting Sgr~A* on the long baseline between the
James Clerk Maxwell Telescope (JCMT) in Hawai`i and the Arizona Radio
Observatory's Submillimeter Telescope (SMT).  Combining this result
with a shorter-baseline detection from the SMT to the Combined Array
for Research in Millimeter-wave Astronomy (CARMA) in California,
Doeleman \etal\ were able to place a size limit of 37~$\mu$as on the
emission if its distribution is a circular Gaussian.  They conclude
that the emission must be partially optically thin and offset from the
center of Sgr~A*, since an optically thick sphere centered on the
black hole would be lensed to a larger size.

With only detections on two baselines, the data are insufficient to
distinguish between other models such as a ``doughnut'' of emission as
might be expected from a face-on accretion disk.  In actuality, the
structure of the emission from Sgr~A* is likely to be much more
complicated due to the inclination of the system and general
relativistic effects, such as Doppler boosting of approaching emission
and weakening of emission from in front of the black hole due to
gravitational redshift.  It is this complexity of source structure
that drives the need for more data.  The eventual goal is to produce
an image of the quiescent emission in Sgr~A*, from which many physical
parameters (such as the spin of the black hole) can be derived.  In
the nearer term, nonimaging methods of data analysis can be used to
produce scientific output from millimeter VLBI observations.

\section{Parameter Estimation}

The \cite[Doeleman \etal\ (2008)]{doeleman08} detections measure
correlated flux densities on certain spatial scales and orientations,
which can be used to constrain parameters in ensembles of models.  For
instance, \cite[Broderick \etal\ (2009)]{broderick09} generate an
ensemble of radiatively inefficient accretion flow (RIAF) models of
disk emission from Sgr~A* meeting certain prior constraints (such as
the multiwavelength spectrum of emission from Sgr~A*) and then use the
\cite[Doeleman \etal\ (2008)]{doeleman08} detections to estimate
probable values of the black hole spin, inclination of the spin axis,
and orientation of the accretion disk on the plane of the sky.  Within
the RIAF ensemble, it is found that low-inclination (i.e., nearly
face-on) models are strongly disfavored.  Prospects are excellent for
observations in the near future, which are likely to include a
telescope in Chile, to be able to place strong constraints on the
orientation of the disk within the RIAF context (\cite[Fish
  \etal\ 2009]{fish09}).  As additional telescopes are added to the
millimeter VLBI observing array, the constraints provided by the
detections may be able to obtain a value for the black hole spin even
before a high-quality image can be created.

It will be necessary to distinguish amongst different models of the
mechanism of emission from Sgr~A* (e.g., jets, fully general
relativistic magnetohydrodynamic simulations, etc.), both in order to
understand the accretion/outflow physics in general and because each
model will have different particular predictions for parameter values
such as the black hole spin.  Thus, there is a need for similar
analyses to be carried out in order to interpret present and future
millimeter VLBI results.

\firstsection
\section{Imaging with an Event Horizon Telescope}

As additional telescopes are added to the millimeter VLBI array, it
will become possible to produce an image of the emission from Sgr~A*.
Current observations have used the SMT, CARMA dishes, and a phased
array of the JCMT, Caltech Submillimeter Observatory, and
Submillimeter Array dishes on Mauna Kea.  In the near future, it may
be possible to extend this array by including the Atacama
Submillimeter Telescope Experiment (ASTE) or Atacama Pathfinder
Experiment (APEX) in Chile, the Large Millimeter Telescope (LMT) under
construction in Mexico, the Institut de Radioastronomie
Millim\'{e}trique (IRAM) 30~m telescope at Pico Veleta in Spain, and
the IRAM Plateau de Bure interferometer in France.

\begin{figure}[ht]
\begin{center}
\includegraphics[width=1.3truein]{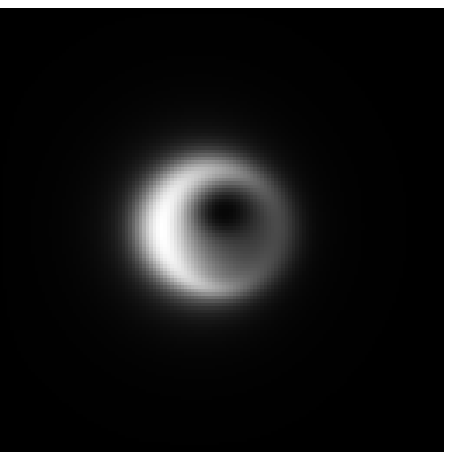}
\includegraphics[width=1.3truein]{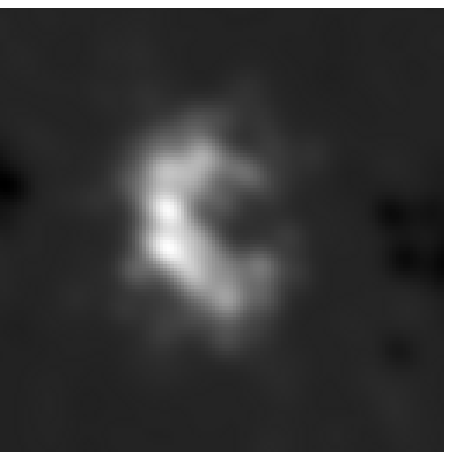}
\includegraphics[width=1.3truein]{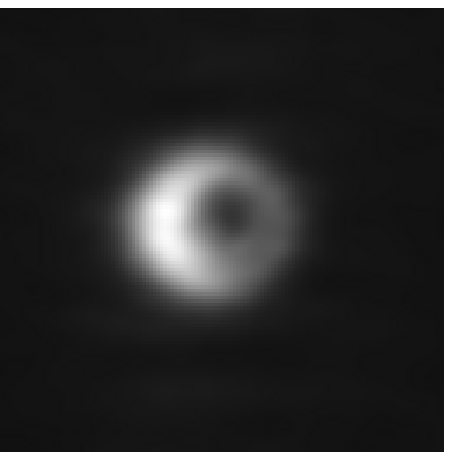}
\end{center}
\caption{\emph{Left:} Model of RIAF emission (345 GHz, $a = 0$, $i =
  30^\circ$), courtesy A.\ Broderick.  \emph{Center:} Simulated image
  from 7-telescope array of the near future.  The black hole shadow is
  easily detected.  \emph{Right:} Simulated image from the
  13-telescope array described in the text.
\label{figure-images}}
\end{figure}

\begin{figure}[ht]
\begin{center}
\includegraphics[width=1.66truein]{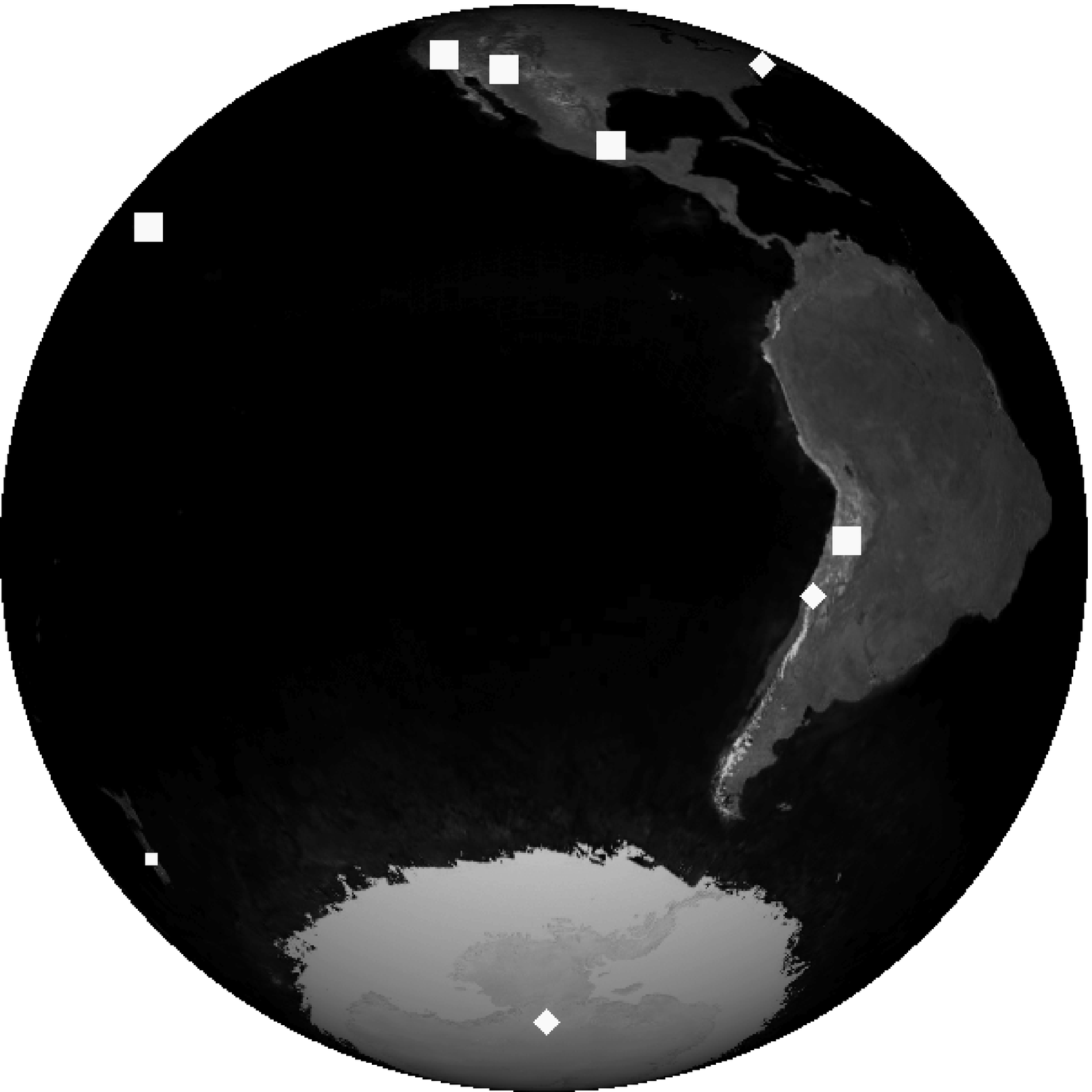}
\includegraphics[width=1.66truein]{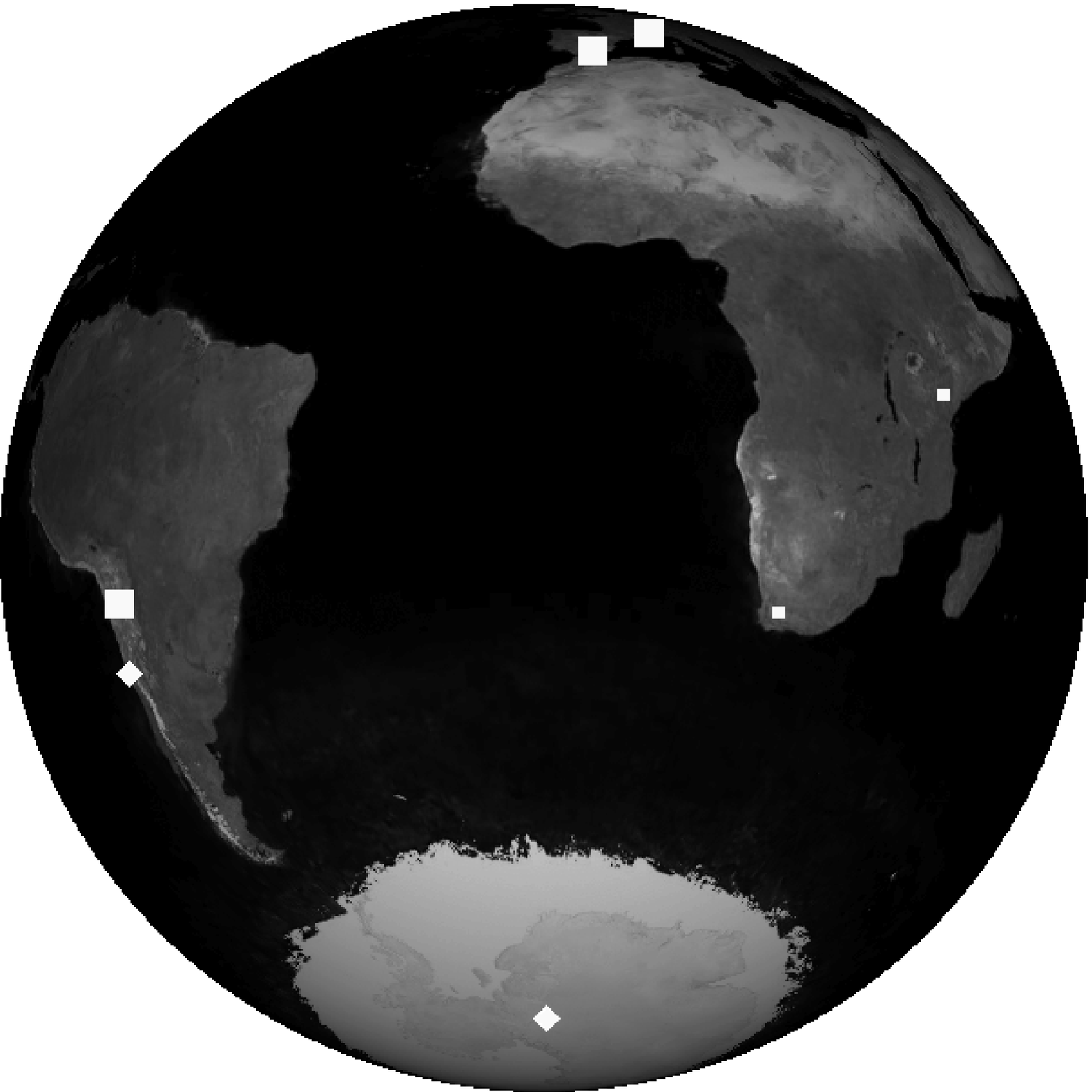}
\includegraphics[width=1.66truein]{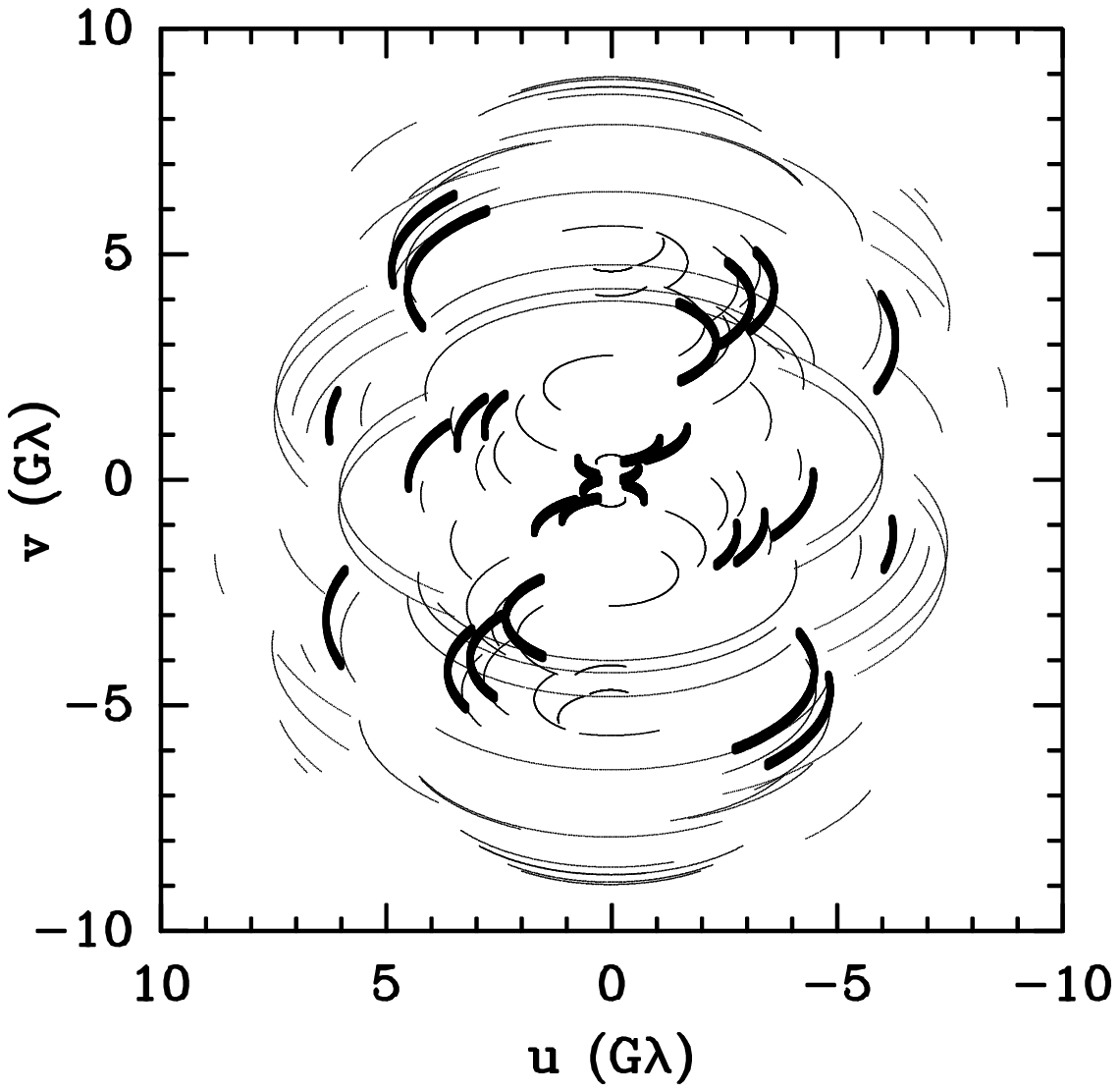}
\end{center}
\caption{\emph{Left two panels:} Current and future millimeter VLBI
  arrays.  Large squares show telescopes that exist or are under
  construction.  Medium diamonds show existing telescopes that could be
  refitted for VLBI.  Small squares show other potential sites.
  \emph{Right:} The $(u,v)$ coverage produced by these arrays at
  230~GHz.  The bold curves show the coverage obtainable from the
  7-telescope array described in the text.  The curves in normal
  weight show the additional coverage that would be provided by the
  13-station Event Horizon Telescope.  High-fidelity imaging requires
  minimizing the gaps in $(u,v)$ coverage.
\label{figure-uv}}
\end{figure}

With upgrades to some of these telescopes (\S \ref{future}), it will
be possible to produce an image that shows the general morphology of
the emission from Sgr~A* (Figure~\ref{figure-images}), possibly
including the ``shadow'' of the black hole (e.g., \cite[Falcke
  \etal\ 2000]{falcke00}).  Unfortunately, image fidelity will be
limited by the placement of these telescopes.  Effectively, VLBI
baselines are sensitive to the Fourier transform of the sky emission,
where the Fourier components are parameterized by $u$ and $v$, the
projected baseline lengths (in wavelengths) as viewed from Sgr~A*.
Longer baselines provide higher angular resolution.  As the Earth
rotates, the projected baselines sweep out arcs in the $(u,v)$ plane,
allowing an image to be produced by an inverse Fourier transform and
deconvolution.  However, the lack of intermediate-spacings in some
directions means that large parts of the $(u,v)$ plane are unsampled.
Reconstructing a high-fidelity, model-independent image of Sgr~A* will
require the addition of a few telescopes to fill in these holes in the
$(u,v)$ plane.  Several existing telescopes, such as the South Pole
Telescope, the Haystack 37~m in Massachusetts, and the Swedish-ESO
Submillimeter Telescope in Chile, could be upgraded for use in a
millimeter VLBI array.  It may also be desirable to place additional
telescopes in geographically favorable locations such as South Africa,
Kenya, and New Zealand (Figure~\ref{figure-uv}).  Expanding the
millimeter VLBI array by upgrading existing telescopes and adding a
few new ones would allow very high quality imaging of Sgr~A*.

\section{Tracking Flaring Structures in Real Time with VLBI}

While very high angular resolution millimeter imaging awaits the
availability of additional VLBI telescopes, millimeter VLBI
visibilities are already producing useful science.  Indeed, the
scientific implications of the \cite[Doeleman
  \etal\ (2008)]{doeleman08} detection stem from measuring the
visibility amplitude on the JCMT-SMT baseline, which is a measure of
the correlated flux density on small spatial scales.  As Sgr~A* is
detected with larger arrays, it will also be possible to produce
closure quantities, good observables that are very sensitive to the
source structure of Sgr~A* but independent of most calibration errors.
The closure phase $\phi_{123} = \phi_{12} + \phi_{23} + \phi_{31}$,
the sum of visibility phases around a closed triangle of three
antennas, is robust against most phase errors, especially those caused
by a variable troposphere.  A nonzero closure phase implies source
structure asymmetry.  Similarly, a closure amplitude $A_{1234} =
|A_{12}||A_{34}|/(|A_{13}||A_{24}|)$, the ratio of visibility
amplitudes at 4 telescopes, is robust against systematic gain
calibration errors.  The number of independent closure quantities
grows quickly with the number of antennas in an array.

\begin{figure}
\begin{center}
\includegraphics[height=2.0truein]{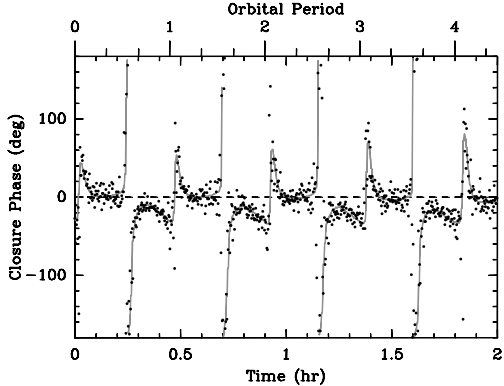}
\end{center}
\caption{Simulated 230~GHz closure phases of a hot spot orbiting at
  the ISCO embedded in a quiescent RIAF disk ($a = 0, i = 30^\circ$).
  The grey curve shows the model values for an array consisting of
  phased Mauna Kea (JCMT + CSO + 6 SMA dishes), phased CARMA (8
  dishes), and the SMT.  Black points represent simulated 10 second
  integrations at 8~Gbit\,s$^{-1}$.
\label{fig-closures}}
\end{figure}

The frequent flaring of Sgr~A* implies that its source structure
changes rapidly.  Simulations by \cite[Doeleman
  \etal\ (2009)]{doeleman09} show that the sensitivity of VLBI
combined with a time resolution much shorter than the natural orbit
period make it very likely that time-variable structures during flares
can be detected over the course of a single night of observing.
Within the next few years, millimeter VLBI will have the sensitivity
to probe $r_\mathrm{Sch}$-scale structural changes on time scales of
10~seconds (Figure~\ref{fig-closures}).  This time resolution will be
critical for detecting source structure periodicity, as might be
expected in orbiting hot spot models (\cite[Broderick \& Loeb
  2005]{broderick05}).  Rapid periodicity may provide strong evidence
that the black hole in Sgr~A* has spin, since the period of the
innermost stable circular orbit (ISCO) decreases rapidly with
increasing spin.  Depending on the spin of the black hole, the ISCO
period may be as long as nearly half an hour or as short as
approximately 4 minutes.

\section{Future Enhancements \label{future}}

Several technological advancements are currently in progress to
increase the sensitivity of the millimeter VLBI array.  A phased array
processor to sum the collecting area on Mauna Kea (\cite[Weintroub
  2008]{weintroub08}) has been tested.  Similar hardware could be used
to increase sensitivity at CARMA, Plateau de Bure, and the Atacama
Large Millimeter/submillimeter Array (ALMA) in Chile.  Digital
backends (DBEs) have been developed to process 1 GHz of data (4
Gbit\,s$^{-1}$ with 2-bit Nyquist sampling), and next-generation DBEs
will improve upon this by a factor of four.  Mark 5B+ recorders can
already record 2 Gbit\,s$^{-1}$ data streams (presently requiring two
at each site per DBE), and the Mark 5C recorders currently being
developed will be able to handle even faster data rates.  Cryogenic
sapphire oscillators are being examined as a possible frequency
standard to supplement or replace hydrogen masers to provide greater
phase stability, which may improve coherence at higher frequencies.

Future observations will initially focus on improving sensitivity by
observing a wider bandwidth and using phased array processors.  Dual
polarization observations will become a priority not only for the
$\sqrt{2}$ improvement in sensitivity for total-power observations but
also to allow full polarimetric VLBI of Sgr~A*.  Higher frequency
observations, such as in the 345~GHz atmospheric window, will provide
even greater spatial resolution in a frequency regime where
interstellar scattering and optical depth effects are minimized.

The timing is right to move forward on building an Event Horizon
Telescope to produce high-fidelity images of Sgr~A* as well as other
scientifically compelling sources, such as M87.  Receivers currently
being produced en masse for ALMA could be procured for other
millimeter VLBI stations, in many cases providing substantial
improvements in sensitivity.  Studies of climate and weather will be
necessary to provide information on the astronomical suitability of
prospective sites for future telescopes, such as those at the present
ALMA Test Facility or additional telescopes constructed specifically
for millimeter VLBI (which would mesh well with present ALMA
construction).  Some existing telescopes will require improvements to
their systems, such as increasing the bandwidth of the intermediate
frequency signal after mixing.  It will also be highly desirable to
install permanent VLBI hardware at all sites to allow turnkey VLBI
observing in order to maximize the efficiency of VLBI observations in
terms of personnel time and transportation costs.

\section{Conclusions}

Millimeter VLBI offers an unparalleled ability to probe the emission
from Sgr~A* at angular scales of a few $r_\mathrm{G}$ and on
timescales of a few seconds.  Current 1.3~mm VLBI observations have
established that the millimeter emission emanates from a compact
region offset from the center of the black hole.  These data are
already being used to constrain key physical parameters (e.g., spin,
inclination, orientation) in models of the emission (e.g., RIAF
models).  Future additions to the VLBI array would allow the
millimeter emission to be imaged directly.  In the meantime, closure
quantity analysis may allow the spin of the black hole to be inferred
from source structure periodicity.  The technical advancements
necessary to realize these goals are already in progress.

\acknowledgments

VLBI work at MIT Haystack Observatory is supported through grants from
the National Science Foundation.  VLBI at (sub)millimeter wavelengths
is made possible through broad international collaborative efforts and
the support of staff and scientists at all participating facilities.


\begin{thebibliography}{}

\bibitem[Broderick \& Loeb (2005)]{broderick05} {Broderick, A.E., \&
  Loeb, A.} 2005, \textit{MNRAS}, 363, 353

\bibitem[Broderick \etal\ (2009)]{broderick09} {Broderick, A.E., Fish,
  V.L., Doeleman, S.S., \& Loeb, A.} 2009, \textit{ApJ}, 697, 45

\bibitem[Doeleman \etal\ (2009)]{doeleman09} {Doeleman, S.S., Fish,
  V.L., Broderick, A.E., Loeb, A., \& Rogers, A.E.E.} 2009,
  \textit{ApJ}, 695, 59

\bibitem[Doeleman \etal\ (2001)]{doeleman01} {Doeleman, S.S., \etal}
  2001, \textit{AJ}, 121, 2610

\bibitem[Doeleman \etal\ (2008)]{doeleman08} {Doeleman, S.S., \etal} 2008,
  \textit{Nature}, 455, 78

\bibitem[Falcke \etal\ (2000)]{falcke00} {Falcke, H., Melia, F., \&
  Agol, E.} 2000, \textit{ApJ} (Letters), 528, L13

\bibitem[Fish \etal\ (2009)]{fish09} {Fish, V.L., Broderick, A.E.,
  Doeleman, S.S., \& Loeb, A.} 2009, \textit{ApJ} (Letters), 692, L14

\bibitem[Reid (2008)]{reid08} {Reid, M.J.} 2008,
  \textit{Internat.\ J.\ Modern Phys.\ D}, in press, arXiv:0808.2624

\bibitem[Weintroub (2008)]{weintroub08} {Weintroub, J.} 2008,
  \textit{J.\ Phys.\ Conf.\ Ser.}, 131, 012047

\end{thebibliography}
\end{document}